\documentclass[epj]{webofc}
\usepackage[utf8]{inputenc}
\usepackage[varg]{txfonts}   
\usepackage{booktabs}
\usepackage{xcolor}
\definecolor{darkred}{rgb}{0.4,0.0,0.0}
\definecolor{darkgreen}{rgb}{0.0,0.4,0.0}
\definecolor{darkblue}{rgb}{0.0,0.0,0.4}
\usepackage[bookmarks,linktocpage,colorlinks,
    linkcolor = darkred,
    urlcolor  = darkblue,
    citecolor = darkgreen]{hyperref}
\usepackage{subfigure}
\wocname{EPJ Web of Conferences}
\woctitle{Lattice2017}
\newcommand{\dif}{\mathrm{d}}
\newcommand*\pFqskip{8mu}
\catcode`,\active
\newcommand*\pFq{\begingroup
        \catcode`\,\active
        \def ,{\mskip\pFqskip\relax}%
        \dopFq
}
\catcode`\,12
\def\dopFq#1#2#3#4#5{%
        {}_{#1}F_{#2}\biggl[\genfrac..{0pt}{}{#3}{#4};#5\biggr]%
        \endgroup
}
\begin{document}
%
\selectlanguage{english}
\title{%
Explicit positive representation for complex weights on $\mathbb R^d$
}
\author{%
\firstname{Błażej} \lastname{Ruba}\inst{1}\fnsep\thanks{Speaker, \email{blazej.ruba@student.uj.edu.pl}}\fnsep\thanks{Acknowledges financial support by the NCN grant: UMO-2016/21/B/ST2/01492.} \and
\firstname{Adam}  \lastname{Wyrzykowski}\inst{1}\fnsep\thanks{Acknowledges financial support by the NCN grant: UMO-2016/21/B/ST2/01492.}
}
\institute{%
Jagiellonian University, Łojasiewicza 11, 30-348 Kraków, Polska
}
\abstract{%
  It is an old idea to replace averages of observables with respect to a complex weight by expectation values with respect to a genuine probability measure on complexified space. This is precisely what one would like to get from complex Langevin simulations. Unfortunately, these fail in many cases of physical interest. We will describe method of deriving positive representations by matching of moments and show simple examples of successful constructions. It will be seen that the problem is greatly underdetermined.
  

}
\maketitle
\section{Introduction}\label{sec:intro}

Numerical simulations involving Monte Carlo methods are only possible if averages with respect to genuine probability distributions are to be computed. This fact prevents lattice calculations in theories defined by complex effective actions, including QCD at finite chemical potential. In \cite{Parisi_complex_prob,Klauder_coh_st_lang} it was proposed that probability distribution could be constructed by setting up a stochastic process on complexification of the integration manifold - the idea known as the complex Langevin approach. It was speculated that the simulation time average of this process converges to the correct value of the path integral, in analogy with the real Langevin dynamics. In \cite{Nakazato,Okano,Aarts_complang_trusted} arguments for this were presented. Unfortunately, in applications \cite{AFC1_CLEproblems,BMS,Wosiek_Haymaker,SinclairProceedings} many difficulties were encountered. In particular convergence to wrong results was occasionally observed. Much effort was put it understanding these phenomena \cite{Salcedo93,Aarts_complang_etiology,Nishimura,Salcedo16,GaugeCooling,SeilerProceedings}, but complete picture is not available until today. Nevertheless, it is known that probabilistic representations of complex measures exist under mild assumptions. Some constructions were presented in \cite{Salcedo97,Weingarten,Salcedo07,Wosiek_zzbar,Salcedo16Gibbs,SalcedoProceedings,WosiekProceedings,WyrzykowskiProceedings}. In \cite{Wosiek_Seiler} representations of a class of measures on compact group manifolds $U(1)^d$ were constructed using Fourier techniques. In this work we extend this method to distributions on real vector spaces $\mathbb R^d$. In Section \ref{sec:mom-match} we explain the general strategy. It is based on directly solving the matching conditions, which express compatibility of probabilistic measure with given complex weight. For simplicity of presentation we start from the case of one degree of freedom. Then we discuss the issue of imposing positivity of constructed distributions. Large underdeterminacy of the problem is exploited by choosing an ansatz for the solution of matching conditions which leads to particularly simple final formula. It turns out that positive measure can be expressed by the Fourier transform of the complex weight continued to complex wave vectors. We show that for weights decreasing at infinity sufficiently rapidly this continuation exists. Moreover our construction defines an analytic function decreasing rapidly at infinity. Subsection \ref{sec:torii} contains a brief discussion of distributions on compact group manifolds. In Section \ref{sec:examples} we discuss examples. Extension to larger number of variables is described in Section \ref{sec:generalize}. We summarize in Section \ref{sec:summary}.

\section{Construction of positive representations}\label{sec:mom-match}

\subsection{Matching conditions} \label{sec:mom-match-sub}

In what follows we assume that a complex weight $\rho$ on $\mathbb R^d$ is given and ask whether it is possible to construct a probability distribution function $P$ on $\mathbb C^d = \mathbb R^{2d}$ such that for some class of analytic functions $\mathcal O$ we have
    \begin{equation}
  \left \langle \mathcal O \right \rangle_{\rho} \equiv  \frac{\int_{\mathbb R^d} \dif^d t \rho (t) \mathcal O(t)}{\int_{\mathbb R^d} \dif^d t \rho (t)} = \int_{\mathbb R^{2d}} \dif^d x \dif^d y P(x,y) \mathcal O(x+iy) \equiv  \left \langle \mathcal O \right \rangle_{P}.
  \label{eq:BCL_condition}
    \end{equation}
In other words, it is required that expectation values of observables with respect to the two distributions agree. First we will consider the case of one degree of freedom and then generalize the results to larger number of variables. Note that it is formally sufficient to consider polynomial observables, i.e. impose equality of moments
\begin{equation}
\left \langle t^k  \right \rangle_{\rho(t)} = \left \langle (x+i y)^k \right \rangle_{P(x,y)}.
\label{eq:mom_match}
\end{equation}
If this condition is satisfied, (\ref{eq:BCL_condition}) follows from Taylor-expanding $\mathcal O(t)$ and applying (\ref{eq:mom_match}) term by term. For this argument to be correct it is required to justify that we can interchange infinite summation involved in Taylor expansion of $\mathcal O(t)$ with integration. This point will be addressed later.

In order to solve the moment matching condition (\ref{eq:mom_match}) we introduce polar coordinates $(r, \theta)$ in $\mathbb R^2 = \{ (x,y) \}$ and perform Fourier transformation in angular direction only:
\begin{equation}
P(r, \theta) = \frac{1}{2 \pi} \sum_{k= - \infty}^{\infty} P_k(r) e^{i k \theta},
\label{eq:Fourier}
\end{equation}
where $P_{-k}(r)=P_k(r)^*$ by assumed reality of $P(r, \theta)$. Equation $(\ref{eq:mom_match})$ is then equivalent to
\begin{equation}
\int _{0}^{\infty} \dif r \ P_{-k}(r) r^{k+1} = \left \langle t^k \right \rangle_{\rho (t)}, \quad k \geq 0.
\end{equation}
Clearly this system is greatly underdetermined - we have a countably infinite family of arbitrary functions $P_k(r)$, each subject to one linear constraint. 

\subsection{Conditions for positivity}

Solving the matching conditions using Fourier transformation introduces additional difficulty - it is in general difficult to decide whether a function given by Fourier series is positive. Abstract characterization is provided by Bochner's theorem \cite{SimonReedII}, but checking whether its assumptions are satisfied is intractable in practice. Therefore it is desirable to find less general, but simpler criteria. 


Our construction is based on the observation that in the sum (\ref{eq:Fourier}) only the lowest ($k=0$) term can be positive everywhere. Therefore we anticipate that it should be the largest one if the whole sum is to be positive. This can be made more precise as follows. For nonnegative $P(r, \theta)$ we have $|P_k(r)| \leq P_0(r)$ for all $k$:
\begin{equation}
|P_k(r)| = \left| \int_0^{2 \pi} \dif \theta e^{-ik \theta} P(r, \theta)    \right| \leq \int_0^{2 \pi} \dif \theta P(r, \theta) = P_0 (r).
\end{equation}
In particular, $P_0(r)$ is nonnegative everywhere. Therefore the triangle inequality implies that
\begin{equation}
P(r, \theta) \geq \frac{1}{2 \pi} P_0(r) - \frac{1}{2 \pi} \sum_{k \neq 0} |P_k(r)|.
\end{equation}
Hence for positivity of $P(r, \theta)$ it is sufficient that we have an estimate
\begin{equation}
\sum_{k= - \infty}^{\infty} |P_k(r)| \leq P_0(r).
\label{eq:positivity_suff}
\end{equation}
In what follows we will impose this condition.

\subsection{Our ansatz}\label{sec:construction}

There is large freedom in choosing the coefficient functions $P_k(r)$. It should be used to simplify the problem. It will be seen later that the following ansatz achieves this goal: 
\begin{subequations}
    \begin{gather}
        P_0(r) = c_0 \exp (- \sigma_0 r^2), \\
        P_k(r) = c_k r^{|k|} \exp ( - \sigma r^2 ), \quad k \neq 0,
    \end{gather}
\end{subequations}
where $0<\sigma_0 <\sigma$ are free parameters and $c_k=c_{-k}^*$ are to be determined from the moment matching condition. This leads to
\begin{equation}
    c_k = \frac{2 \sigma^{k+1}}{k!} \left \langle t^k  \right \rangle_{\rho(t)}^*, \quad k \geq 0.    
\end{equation}
Distribution $P(r, \theta)$ takes the form
\begin{equation}
P(r, \theta) = \frac{\sigma_0}{\pi} e^{- \sigma_0 r^2} + \frac{2 \sigma}{\pi} e^{- \sigma r^2} \sum_{k=1}^{\infty} \mathrm{Re} \left \langle \frac{1}{k!} \left( \sigma r e^{- i \theta} t  \right)^k \right \rangle_{\rho(t)}.
\label{eq:P_def_moments}
\end{equation}
Now suppose that it is correct to interchange the infinite summation over $k$ with averaging with respect to $\rho(t)$. Then the summation can be performed for any $\rho(t)$, leading to
\begin{equation}
    P(r, \theta) = \frac{\sigma_0}{\pi} e^{- \sigma_0 r^2} + \frac{2 \sigma}{\pi} e^{- \sigma r^2} \mathrm{Re} \ \left \langle \exp \left( \sigma r e^{- i \theta} t  \right) -1 \right \rangle_{\rho(t)}.
    \label{eq:Gaussian_transform}
\end{equation}
Probability distribution $P$ is expressed in terms of the Fourier transform of $\rho$ extended to complex arguments.

We have been rather cavalier about mathematical correctness of manipulation performed so far. We will now work out a set of conditions for which they are under control. First note that the average with respect to $\rho$ in (\ref{eq:Gaussian_transform}) is unambiguously defined only if $\rho$ decreases fast at infinity. It is sufficient if $|\rho(t)| \leq C e^{- \alpha |t|^{1+ \epsilon}}$ for some positive constants $C, \alpha, \epsilon$. Under this assumption the integral is sufficiently well behaved that it is correct to differentiate under the integral sign, hence we see explicitly that $P$ is an analytic function of $x,y$. We still need to choose $\sigma, \sigma_0$ in such a way that $P$ decreases at infinity and is positive everywhere. We will do this assuming that $\epsilon \geq 1$. Moreover we take (without loss of generality) $Z= \int \dif t \rho(t) =1$. Then moments of $\rho$ can be estimated by
\begin{equation}
\left| \langle t^k \rangle_{\rho(t)} \right| \leq \frac{C \Gamma \left( \frac{k+1}{2} \right)}{\alpha^{k+1}}, 
\label{eq:mom_estimate}
\end{equation}
hence for $k > 0$ we have
\begin{equation}
\frac{|P_k(r)|}{P_0(r)} \leq \frac{C}{\alpha} \frac{\Gamma \left( \frac{k+1}{2} \right)}{k!} \left( \frac{\sigma}{\sigma_0} \right) \left( \frac{\sigma r}{\alpha} \right)^k e^{-(\sigma - \sigma_0)r^2}.
\end{equation}
For $\sigma > \sigma_0$ the right hand side is a bounded function of $r$. Replacing it by its maximum value we obtain
\begin{equation}
\frac{|P_k(r)|}{P_0(r)} \leq \frac{C}{\alpha} \frac{\Gamma \left( \frac{k+1}{2} \right)}{k!} \left( \frac{\sigma}{\sigma_0} \right) \left( \frac{\sigma^2 k}{2e \alpha^2 (\sigma- \sigma_0)} \right)^{\frac{k}{2}}.
\end{equation}
From this inequality and properties of the Gamma function it follows that
\begin{equation}
\sum_{k \neq 0} \frac{|P_k(r)|}{P_0(r)} \leq \frac{4 \sqrt{\pi} C}{\alpha} \frac{\sigma}{\sigma_0} \sum_{k=1}^{\infty} \frac{k^{\frac{k}{2}}}{\Gamma \left( \frac{k+2}{2} \right)} \left( \frac{\sigma}{8 \alpha^2 e (1 - \frac{\sigma_0}{\sigma})} \right)^{\frac{k}{2}}.
\end{equation}
For fixed $\frac{\sigma_0}{\sigma}<1$ and sufficiently small $\sigma$ the series on the right hand side converges absolutely. Hence it defines an analytic function of $\sqrt{\sigma}$ with Taylor expansion starting at the linear term. Therefore it can be made as small as one wishes if sufficiently small $\sigma$ is chosen. Criteria derived in previous section show that this is sufficient for positivity of $P$. Moreover we see that there exist positive constants $0<a<A$ such that
\begin{equation}
a P_0(r) \leq  P(r, \theta)  \leq A P_0(r).
\end{equation}
Estimates for $|P(r, \theta)|$ and $|P_k(r, \theta)|$ derived so far are sufficiently strong to justify the interchange of sums and integrals performed to derive (\ref{eq:Gaussian_transform}). Hence (\ref{eq:mom_match}) is satisfied. Moreover for functions of the form
\begin{equation}
\mathcal O (t) = \sum_{k=0}^{\infty} \mathcal O_k t^k 
\end{equation}
we see from (\ref{eq:mom_estimate}) that for correctness of interchanging the order of summation over $k$ and averaging with respect to $\rho$ and $P$ (and hence for (\ref{eq:BCL_condition})) it is sufficient that the coefficients $\mathcal O_k$ satisfy
\begin{equation}
\sum_{k=0}^{\infty} \frac{\Gamma \left( \frac{k+1}{2} \right) |\mathcal O_k|}{\alpha^k} < \infty.
\end{equation}

\subsection{Distributions on compact group manifolds}\label{sec:torii}

The reasoning leading to (\ref{eq:Gaussian_transform}) can be performed also for distributions $\rho( \vec t)$ on torii $U(1)^d$. In this case a probabilistic measure $P(x,y)$ on complexified manifold $U(1)^d \times \mathbb R^d$ can be represented as
\begin{equation}
P(\vec x, \vec y) = \frac{1}{(2 \pi)^d} \sum_{\vec k \in \mathbb Z^d} P_{\vec k} (\vec y) e^{ i \vec k \cdot \vec x},
\end{equation} 
where the coefficient functions $P_{\vec n}(\vec y)=P_{-\vec n}(\vec y)^*$ satisfy the moment matching conditions
\begin{equation}
\int_{\mathbb R^d} \dif^d y P_{ \vec k} (\vec y) e^{ \vec k \cdot \vec y} = \left \langle e^{-i \vec k \cdot \vec t} \right \rangle_{\rho (\vec t)}.   
\end{equation}
Several proposals for $P_{\vec n}(\vec y)$ were explored in \cite{Wosiek_Seiler}. In particular it was found that $P(\vec x, \vec y)$ can be chosen as a convolution of weight $\rho(\vec t)$ with a smoothing kernel on $U(1)^d$. Generalization to nonabelian compact groups is also possible, see e.g. \cite{Salcedo07}.

\section{Examples}\label{sec:examples}

\subsection{Gaussian weights}

\begin{figure}
    \centering
    \includegraphics[width=8cm]{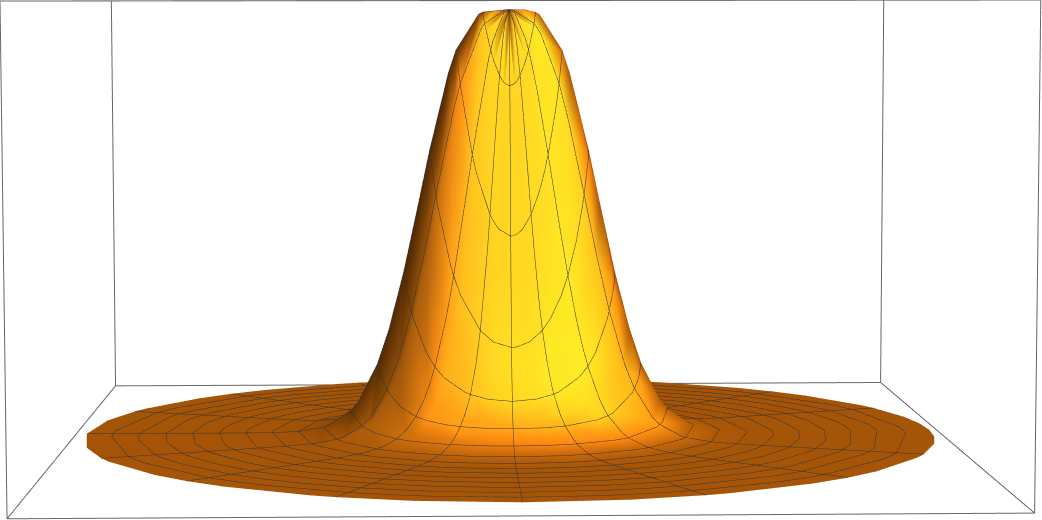}
    \caption{Positive representation of a complex gaussian.}
    \label{fig:gaussian_positive_rep}
\end{figure}

As a first example we choose complex gaussians $\rho(t)= e^{- \lambda t^2}$, $\lambda = \lambda_R + i \lambda_I$. Then (\ref{eq:Gaussian_transform}) leads to
\begin{equation}
P(r, \theta) = \frac{\sigma_0}{\pi} e^{- \sigma_0 r^2} + \frac{2\sigma}{\pi} e^{- \sigma r^2} \mathrm{Re} \left( \exp \left( \frac{ \sigma^2 r^2 e^{2 i \theta}}{4 \lambda} \right) -1 \right).
\end{equation}
Plot of $P(r, \theta)$ for $\lambda=\frac{1+i}{2}$, $\sigma_0= \frac{1}{4}$, $\sigma= \frac{2}{5}$ is presented in Fig. \ref{fig:gaussian_positive_rep}. The limit $\mathrm{Arg} \lambda \to \pm \frac{\pi}{2}$ converges and moments of $\rho$ are properly reproduced also in this case. This is in contrast with the positive representation found in \cite{AmbjornYang} using the complex Langevin method,
\begin{equation}
P(r, \theta) = N e^{-2 (1+q^2) \lambda_R r^2} \sum_{n= - \infty}^{\infty} e^{i n (2 \theta + \psi)} I_n \left( \frac{2 q \lambda_R r^2}{\sin \psi} \right) = N e^{- 2 \lambda_R \left( x^2 + 2qxy + (1+2q^2) y^2 \right) } ,
\end{equation}
where $q=\cot \psi = \frac{\lambda_R}{\lambda_I}$, $N=\frac{2 \lambda_R \sqrt{1+q^2}}{\pi}$. This distribution is an example of positive representation not satisfying (\ref{eq:positivity_suff}).

\subsection{Exponential of a monomial}

\begin{figure}
    \centering
    \includegraphics[width=8cm]{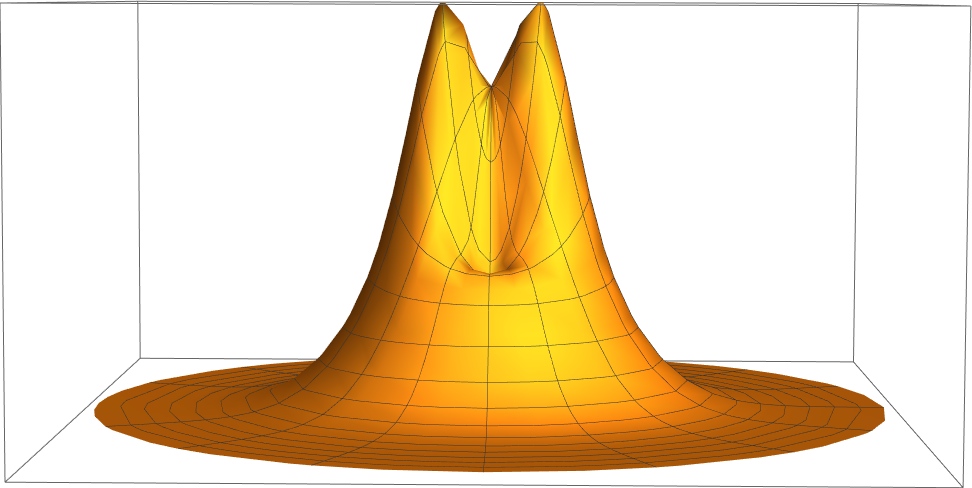}
    \caption{Positive representation of an exponential of a quartic monomial.}
    \label{fig:quartic_positive_rep}
\end{figure}

Now we consider $\rho(t)= e^{- \lambda t^4}$. Equation (\ref{eq:Gaussian_transform}) gives
\begin{equation}
P(r, \theta) =  \frac{\sigma_0}{\pi} e^{- \sigma_0 r^2} + \frac{2 \sigma}{\pi} e^{- \sigma r^2} \mathrm{Re} \left( \pFq{0}{2}{}{\frac{1}{2}, \frac{3}{4}}{\frac{\zeta^2}{256}} -1 +
\frac{\Gamma \left( \frac{3}{4} \right) \zeta}{2 \Gamma \left( \frac{1}{4} \right) }  \pFq{0}{2}{}{\frac{5}{4}, \frac{3}{2}}{\frac{\zeta^2}{256}} \right),
\end{equation}
where $\zeta = \frac{\sigma^2 r^2 e^{2i \theta}}{\sqrt{\lambda}}$ and $\pFq{0}{2}{}{b_1,b_2}{z}$ is the generalized hypergeometric function:
\begin{equation}
\pFq{p}{q}{a_1,...,a_p}{b_1,...,b_q}{z}=\sum_{k=0}^{\infty} \frac{a_1(a_1+1)...(a_1+k-1)...a_p(a_p+1)...(a_p+k-1)}{b_1(b_1+1)...(b_1+k-1)...b_q(b_q+1)...(b_q+k-1)} \frac{z^k}{k!}.
\end{equation}
Distribution $P(r, \theta)$ is presented in Fig. \ref{fig:quartic_positive_rep}. This calculation extends without further difficulties to exponentials of higher degree monomials, the only difference being that in this case higher hypergeometric functions $\pFq{p}{q}{a_1,...,a_p}{b_1,...,b_q}{z},$ $p>0$, $q>2$ are needed to express $P(r, \theta)$.

It is interesting to see what happens for the weight $\rho(t) = e^{\frac{it^3}{3}}$. Its moments are defined using regularization: we multiply $\rho(t)$ by a test function $\phi (\epsilon t)$ such that $\phi(0)=1$. It is required that $\phi$ is smooth and vanishes at infinity together with all its derivatives faster than any power of $\frac{1}{t}$. Then we take the limit $\epsilon \to 0$. The result is finite and doesn't depend on the choice of $\phi$. This can be seen by repeatedly using the identity $\frac{- i e^{-it}}{t^2+1} \frac{\mathrm{d}}{\mathrm{d} t} e^{it} \rho(t) = \rho(t)$ and integrating by parts. Each steps makes the integrand better behaved at infinity due to powers of $\frac{1}{t^2+1}$. After this is done sufficiently many times the limit $\epsilon \to 0$ can be taken under the integral and dependence on $\phi$ disappears. The average in (\ref{eq:Gaussian_transform}) is divergent for $\theta \neq \pm \frac{\pi}{2}$. Analytic continuation from this line to whole plane yields
\begin{equation}
P(r, \theta) = \frac{\sigma_0}{\pi} e^{- \sigma_0 r^2} + \frac{\sigma}{\pi \mathrm{Ai}(0)} e^{- \sigma r^2} \left( \mathrm{Ai} \left( i \sigma r e^{i \theta} \right) + \mathrm{Ai} \left( -i \sigma r e^{-i \theta} \right) - 2 \mathrm{Ai}(0) \right),
\end{equation}
where $\mathrm{Ai}$ is the Airy function and $\mathrm{Ai}(0) = \frac{1}{3^{\frac{2}{3}} \Gamma \left( \frac{2}{3} \right)}$ is its value at $0$. Direct computation shows that the moment matching condition (\ref{eq:mom_match}) is satisfied. Moreover the procedure described above turns out to be equivalent to deforming the contour of integration in complex $t$ plane to make $\rho(t)$ exponentially decreasing for $t \to \infty$. In this example formula (\ref{eq:P_def_moments}) holds, but the interchange of order of summation and integration leading to (\ref{eq:Gaussian_transform}) is incorrect.

\subsection{Exponential of a polynomial}

\begin{figure}
    \centering
    \includegraphics[width=8cm]{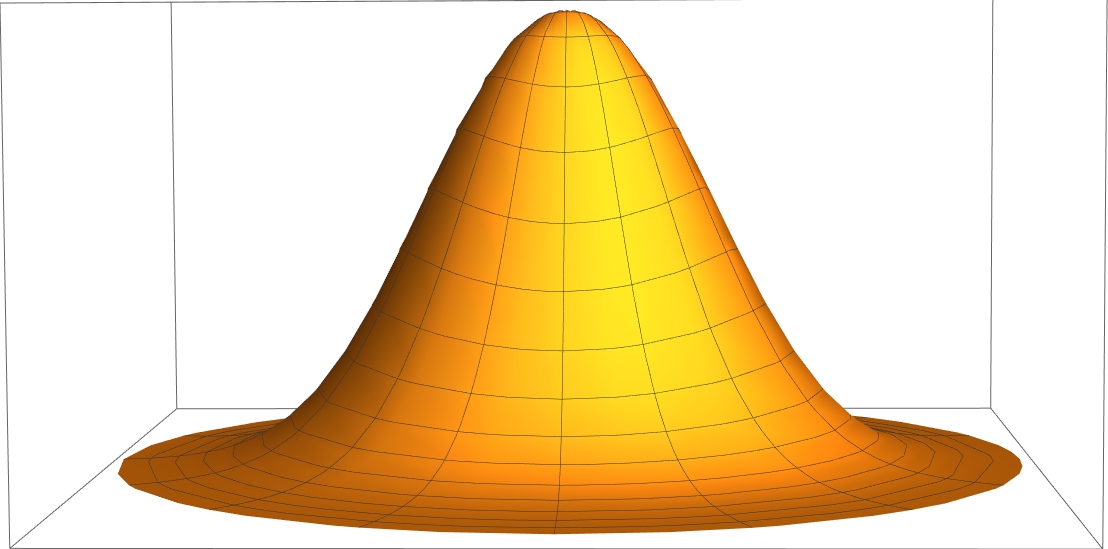}
    \caption{Positive representation of an exponential of $\rho(t)=\exp \left( -i t - \frac{1}{2} t^2 + i t^3 - t^4 \right)$.}
    \label{fig:polynomial_positive_rep}
\end{figure}

For exponential of a general polynomial analytic computation of the average in (\ref{eq:Gaussian_transform}) turns out to be intractable. Figure (\ref{fig:polynomial_positive_rep}) presents the results of numerical integration for $\rho(t)=\exp \left( -i t - \frac{1}{2} t^2 + i t^3 - t^4 \right)$. The coefficients of the polynomial in the exponent were chosen just for the sake of example. Quite general oscillatory weights can be treated in this manner.

\section{Extension to larger number of variables}\label{sec:generalize}

Extension of (\ref{eq:Gaussian_transform}) to larger number of variables is straightforward. Extended space $\mathbb R^{2d}$ is parametrized by $d$ radii and $d$ angles ($x_1+iy_1 = r_1 e^{i \theta_1}$, $x_2+iy_2 = r_2 e^{i \theta_2}$ and so on). Then we can take $P$ to be
\begin{equation}
    P \left(\vec r, \vec \theta \right) = \left( \frac{\sigma_0}{\pi} \right)^d e^{- \sigma_0  r^2} + 2 \left( \frac{\sigma}{\pi} \right)^d e^{- \sigma r^2} \mathrm{Re} \left \langle \exp \left( \sigma \vec t \cdot \vec z^*  \right) -1 \right \rangle_{\rho (\vec t)},
    \label{eq:gaussian_transform_multidim}
\end{equation}
where $\vec z^*=(r_1 e^{-i \theta_1},...,r_d e^{-i \theta_d})$. Arguments presented in Section \ref{sec:construction} can be repeated with only minor modifications to account for the larger number of variables. Therefore one can choose coefficients $\sigma$, $\sigma_0$ so that (\ref{eq:gaussian_transform_multidim}) defines an analytic probability distribution function on $\mathbb R^{2d}$, at least if $\rho$ vanishes rapidly at infinity.

\subsection{Gaussian distributions}

For Gaussian distributions $\rho$
\begin{equation}
    \rho (\vec t) = \exp \left( - \frac{1}{2} t_i M_{ij} t_j, \right)
\end{equation}
where $M$ is a symmetric matrix with all eigenvalues in the right half-plane the formula (\ref{eq:gaussian_transform_multidim}) gives
\begin{equation}
P(r, \theta) = \left( \frac{\sigma_0}{\pi} \right)^d e^{- \sigma_0 r^2} + 2 \left( \frac{\sigma}{\pi} \right)^d e^{- \sigma r^2} \mathrm{Re} \left[ \exp \left( \frac{\sigma^2}{2} z_i^* M^{-1}_{ij} z_j^* \right) -1 \right].
\end{equation}
To avoid evaluation of $M^{-1}$ one may change variables to $w_i^* = M_{ij}^{-1} z_i^*$, $w_i = (M_{ij}^{-1})^* z_i$ at the cost of introducing positive Jacobian $|\mathrm{det} \ M|^2$.

\section{Summary}\label{sec:summary}

We have constructed probabilistic representations of complex weights on $\mathbb R^d$ by explicitly solving the moment matching conditions. Simple Gaussian ansatz led to expression in terms of the weight's Fourier transform extended to complex wave vectors. This relation is remarkably simple. Its mathematically rigorous justification required restriction to weights which decay rapidly at infinity. Nevertheless, we successfully applied it also to pure phase measures. Closed-form expressions were obtained for Gaussian distributions and for exponentials of monomials. Representation for exponential of a general polynomial was obtained in a particular case numerically. The most important challenge for future is to find out whether this approach can be applied efficiently for distributions depending on large numbers of variables.


\clearpage
\bibliography{Lattice2017_352_RUBA}

\end{document}